\documentclass[aps,prc,twocolumn,floatfix,superscriptaddress]{revtex4}
\usepackage{epsfig}
\usepackage{amsmath}
\usepackage{color}

\usepackage{graphicx}

\begin{document}

\title{Correlation between initial spatial anisotropy and final momentum anisotropies in relativistic heavy ion collisions }
\author{Sanchari Thakur}
\email{s.thakur@vecc.gov.in}
\affiliation{Variable Energy Cyclotron Centre, HBNI, 1/AF, Bidhan Nagar, Kolkata-700064, India}

\author{Sumit Kumar Saha}
\email{sk.saha@vecc.gov.in}
\affiliation{Variable Energy Cyclotron Centre, HBNI, 1/AF, Bidhan Nagar, Kolkata-700064, India}

\author{Pingal Dasgupta}
\email{pingaldg@fudan.edu.cn}
\affiliation{Key Laboratory of Nuclear Physics and Ion-beam Application (MOE), Institute of Modern Physics, Fudan University, Shanghai 200433, China}

\author{Rupa Chatterjee}
\email{rupa@vecc.gov.in}
\affiliation{Variable Energy Cyclotron Centre, HBNI, 1/AF, Bidhan Nagar, Kolkata-700064, India}

\author{Subhasis Chattopadhyay}
\email{sub@vecc.gov.in}
\affiliation{Variable Energy Cyclotron Centre, HBNI, 1/AF, Bidhan Nagar, Kolkata-700064, India}

\begin{abstract}
The particle momentum anisotropy ($v_n$) produced in relativistic nuclear collisions is considered to be a response of the  initial geometry or the spatial anisotropy $\epsilon_n$ of the system formed in these collisions. The  linear correlation   between  $\epsilon_n$ and $v_n$ quantifies the  efficiency at which the initial spatial eccentricity is converted to final momentum anisotropy in heavy ion collisions. We study the transverse momentum, collision centrality, and beam energy dependence of this  correlation  for  different charged  particles using a hydrodynamical model framework. The ($\epsilon_n -v_n$) correlation is found to be stronger for central collisions and also for n=2 compared to that for n=3 as expected. 
However, the transverse momentum ($p_T$) dependent correlation coefficient shows interesting features which strongly depends on the mass as well as $p_T$ of the emitted particle. The correlation strength is found to be larger for lighter particles in the lower $p_T$ region. 
 We see that the relative fluctuation in anisotropic flow depends strongly in the value of $\eta/s$ specially  in the region $p_T <1$ GeV unlike the correlation coefficient which does not show significant dependence on $\eta/s$. 
\end{abstract}

\maketitle

\section{Introduction} 
The anisotropic flow of hadrons is known as one of the key observables produced in relativistic heavy ion collisions that provides a strong indication of the formation of hot and dense Quark Gluon Plasma (QGP) phase  and its collective behaviour~\cite{greiner,uli, uli1, bhalerao}. The spatial asymmetry in the initial energy density distribution on the overlapping zone between two colliding nuclei gives rise to anisotropic flow where the magnitude of the flow parameters depend on several factors such as particle mass, beam energy, collision centrality, transverse momentum. 

 It is well known that the relativistic hydrodynamics is  one of the most successful model frameworks which has been used extensively to study the evolution of the QGP medium in order to estimate the several final state observables~\cite{uli, hydro1, h1, h2, h3, h4, h5, greiner,ollitrault2,jaiswal, liyan}. 
The simultaneous explanation of the experimental data of the elliptic flow and the charged particle spectra  by hydrodynamical model calculations at RHIC energy was one of the initial milestones  in this field of research which confirms an early thermalization and collective behaviour of the system produced in relativistic heavy ion collisions~\cite{kolb}.

The initial spatial anisotropy ($\epsilon_n$), specially the ellipticity increases significantly from central to mid-central collisions and consequently the magnitude of the elliptic flow coefficient increases towards peripheral collisions. On the other hand, the rise in initial spatial triangularity  ($\epsilon_3$) with collision centrality is relatively  slower compared to that of $\epsilon_2$. 
The efficiency of conversion of the initial spatial eccentricity to the final momentum anisotropy  depends on the  initial state as well as on the evolution of the produced hot and dense matter. Hydrodynamic model calculation can be quite useful to know the initial states (obtained  by tuning the model parameter to reproduce the experimental data) and also the space time evolution as we cannot get direct information of the initial state from experimental data.

The relation between the initial spatial anisotropy and the final state anisotropic flow parameters has been studied by several groups earlier~\cite{greiner,ollitrault, jaiswal, bass, Niemi, bedang, akc, chaina, raju, greco,alver4,nh1,nh2,nh3,nh4,nh6,nh7}. The effect of $\eta$/s as well as of the fractional contributions of the  number of participants ($N_{\rm part}$) and the number of binary collisions ($N_{\rm coll}$) to the initial entropy and/or energy density production was studied for the first time in an interesting work by Niemi {\it {et. al.}}~\cite{Niemi} for Au+Au collisions at RHIC energy. The initial state anisotropies and their uncertainties in ultrarelativistic heavy ion collisions  have been studied from the Monte Carlo Glauber model by Alvioli {\it {et. al.}}~\cite{alvioli}. Recent experimental data have  shown a correlation between the mean transverse momentum of outgoing particles and the anisotropic flow parameter in Pb+Pb collisions at LHC~\cite{atlas}. A theory calculation shows that the magnitude of this correlation can be directly predicted from the initial conditions using the spatial anisotropy $\epsilon_n$~\cite{ollitrault1}. The correlation between the transverse momentum and anisotropic flow parameter using hydrodynamical model framework has  been studied by Bozek {\it {et. al.}}~\cite{bozek, bozek1}.

These studies suggest that the particle transverse momentum, beam energy, collision centrality, all play crucial role in determining the final momentum anisotropy. Thus,  in order to understand the correlation between $\epsilon_n$ and the anisotropic flow better, it is important to know the simultaneous effect of all these parameters in detail.

In this work we study the (linear) correlation between initial $\epsilon_n$ and  the final momentum anisotropies ($v_n$) of positively charged hadrons using a state-of-the art hydrodynamical model calculation. We focus on the elliptic and triangular flow parameters and the corresponding initial eccentricities are obtained from a sufficiently large number of events. It has already been shown in earlier studies that the correlation between $\epsilon_4$ and $v_4$ is significantly weak~\cite{Niemi} and as a result we do not consider this and other higher order harmonics for the present study. The dependence of the correlation coefficient on collision centrality and transverse momentum is studied in detail for three different types of hadrons. We consider Pb+Pb collisions at 2.76A TeV at LHC and Cu+Cu collisions at 200A GeV at RHIC  to study the dependence of correlation strength on the beam energy and  system size. The correlation coefficients and the relative fluctuations in the anisotropic flow parameters are calculated for two different $\eta/s$ values to check the sensitivity of the results to the shear viscosity coefficient. Additionally, we calculate the Normalized Symmetric Cumulant (NSC) between ($v_2$, $v_3$) at different centrality bins for positively charged pion, kaon and protons.


In the next section we briefly discuss the model framework and the initial state produced in heavy ion collisions. We calculate the $p_T$ integrated and $p_T$ dependent correlation coefficients between $\epsilon_n - v_n$ and discuss our results from Pb+Pb collisions at LHC energy in section III and section IV respectively. The correlation coefficients for a small system like Cu+Cu collisions at RHIC energy are studied in the next section in order to understand the system size and beam energy dependence. We show the $\eta/s$ dependence of the correlation coefficient and the relative fluctuations in section VI and the NSC results in section VII. In section VIII we give the summary and conclusions of all the results.

\begin{figure}
\centerline{\includegraphics*[width=8.0 cm,clip=true]{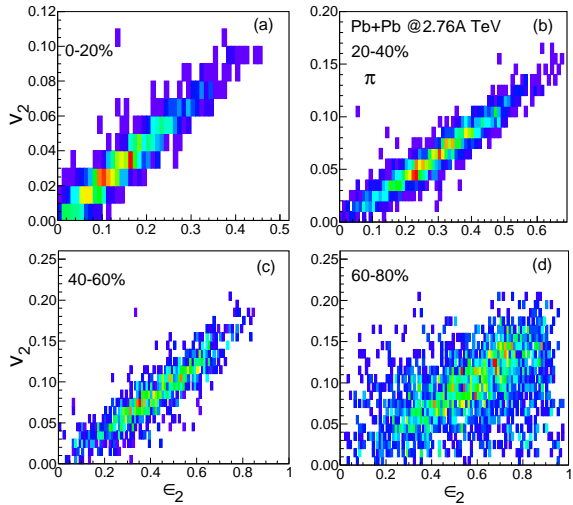}}
\caption{(Color online) Distribution of $\epsilon_2$ and $\pi^+$  $v_2$ at 2.76A TeV Pb+Pb collisions at four different centrality bins. }
\label{pi_v2}
\end{figure}

\begin{figure}
\centerline{\includegraphics*[width=8.0 cm,clip=true]{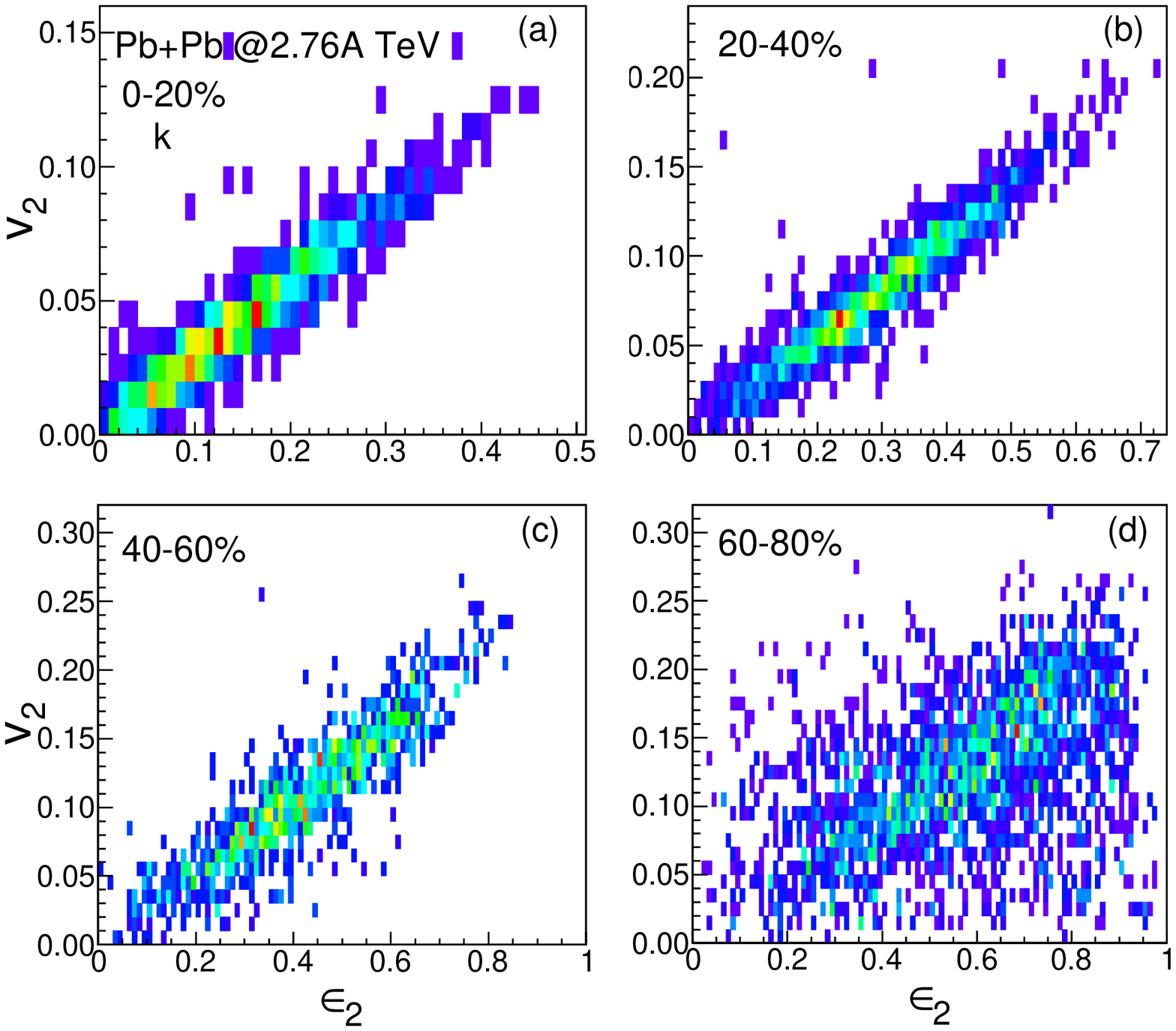}}
\caption{(Color online) Distribution of $\epsilon_2$ and $K^+$  $v_2$ at 2.76A TeV Pb+Pb collisions at four different centrality bins. }
\label{k_v2}
\end{figure}

\begin{figure}
\centerline{\includegraphics*[width=8.0 cm,clip=true]{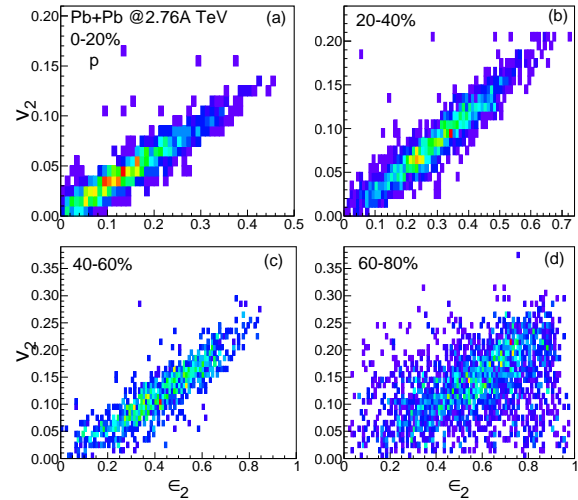}}
\caption{(Color online) Distribution of $\epsilon_2$ and $p$  $v_2$ at 2.76A TeV Pb+Pb collisions at four different centrality bins.}
\label{p_v2}
\end{figure}

\section{Framework}
We use the (2+1) dimensional longitudinally boost invariant hydrodynamical model framework MUSIC~\cite{music} with fluctuating initial conditions to calculate the initial spatial anisotropies and the corresponding anisotropic flow parameters from heavy ion  collisions at different centrality bins  at mid-rapidity. The initial formation time of the plasma is considered as 0.4 fm/$c$ at 2.76A TeV Pb+Pb collisions at the LHC energy.

 A Monte Carlo Glauber initial condition is considered and the value of $\eta/s$ is kept fixed at 0.08 (later we change this value to check the sensitivity of the results to $\eta/s$). 
 The initial energy density is considered to be dependent on a linear combination of soft ($N_{\rm {part}}$) and hard ($N_{\rm {coll}}$) contributions with appropriate weight factors. A constant temperature freeze out is considered and a lattice based equation of state is used for a cross-over transition between the QGP and hadronic matter phases~\cite{lattice}. The centrality bins are selected using an impact parameter range from Ref~\cite{Pb-Pb-bvalue}. The model parameters are set by simultaneously reproducing the experimental data of final state charged particle multiplicity, $p_T$ spectra and anisotropic flow parameters~\cite{Cu-Cu-spectra,Cu-Cu-mult,Cu-Cu-v2,Pb-Pb-spec}. The standard Cooper-Frye formula is used to estimate the production of the hadrons from freeze-out surface~\cite{cooper}.

The initial spatial eccentricity is calculated using the relation \cite{Niemi}:

\begin{equation}
  \epsilon_{n} = -\frac{\int \mathrm{d} x \mathrm{d} y \; r^{n} \cos \left[
  n\left( \phi -\psi_{n}\right) \right] \varepsilon \left(
  x,y,\tau_{0}\right) } {\int \mathrm{d} x \mathrm{d} y \; r^{n} \varepsilon
  \left( x,y,\tau _{0}\right) } \, .
\end{equation}
Where $\psi_{n}$ is the $n^{\rm {th}}$ order event plane angle. 

The corresponding anisotropic flow parameters $v_n$ can be obtained \cite{Niemi} from the invariant particle momentum distribution as :
\begin{equation}\label{eq: v2}
 \frac{dN}{d^2p_TdY} = \frac{1}{2\pi} \frac{dN}{ p_T dp_T dY}[1+ 2\, \sum_{n=1}^{\infty} v_n (p_T) \, \rm{cos} \, n (\phi - \psi_n)] \, .
\end{equation}


\section{Pb+Pb collisions at the LHC}
The  distributions of initial spatial eccentricity ($\epsilon_2$) and the corresponding ($p_T$ integrated) elliptic flow coefficient $v_2$ of positively charged pions for Pb+Pb collisions at $\sqrt{s_{NN}}$ =2.76 TeV at LHC are shown in Fig. 1. The results from hydrodynamical model calculation at mid-rapidity are plotted for centrality bins 0--20\%, 20--40\%, 40--60\%, and 60--80\%. The total number of events used are 2000 and 2400 for 0-20\% and 20-40\% respectively which increase towards more peripheral collisions thereby reducing the statistical errors. The average $\epsilon_2 \ (\epsilon_3)$ values   for 0--20\%, 20--40\%, 40--60\%, and 60--80\% central Pb+Pb collisions are  0.139 (0.099), 0.280 (0.167), 0.405 (0.245), and 0.527 (0.395) respectively. The $\langle \epsilon_3 \rangle$ is found to be about 40\% smaller than $\langle \epsilon_2 \rangle$ for all four centrality bins.

It is to be noted that we consider results upto 80\% centrality bin as the hydrodynamic model calculations for ultra peripheral collisions (more than 80\%) may not give reliable results. The Figs. 2 and 3 show similar $\epsilon_2-v_2$ distribution for positively charged kaons and protons respectively for 4 different centrality bins.

One observes a strong positive linear correlation between pion $v_2$ and $\epsilon_2$  for  all centrality bins and the strength of correlation reduces towards peripheral collisions.
The $\epsilon_n - v_n$ distribution for 60--80\% collision centrality clearly shows that the correlation strength reduces significantly inspite of higher $v_2$ and $\epsilon_2$ coefficients for that centrality bin. 

A similar trend is observed for kaon and protons as well where the correlation  is found to be stronger for more central collisions. 

An estimation of the strength of the linear correlation between two variables is  obtained by dividing the covariance of the variables by the product of their respective standard deviations.
Thus, the correlation coefficient C between the initial spatial eccentricity and final momentum anisotropies can be quantified using the relation \cite{Niemi}:
\begin{equation}
C(\epsilon_n, \, v_n) \, = \, \left \langle \frac{(\epsilon_n - \langle \epsilon_n\rangle_{\rm {av}}) (v_n - \langle v_n \rangle_{\rm {av}})}{\sigma_{\epsilon_n} \sigma_{v_n}}  \right \rangle_{\rm {av}} \, . 
\end{equation}

The quantities  $\sigma_{\epsilon_n}$ and $\sigma_{v_n}$ are the standard deviations of $\epsilon_n$ and $v_n$ respectively. The average is taken using hadron multiplicity  as weight factor. The  correlation coefficient can take any value between -1 to +1. Two quantities are strongly linearly (anti-linearly) correlated when the coefficient is close to 1 (-1). On the other hand, the value of C close to zero implies that the quantities are not correlated linearly.

The correlation coefficients for pion, kaon and proton from Pb+Pb collisions at different centrality bins are shown in Table I. 

The values of the correlation coefficient for pions for 0--20\% and  20--40\% centrality bins remains almost same at about 0.95 (See table 1). For 40--60\% centrality bin, we observe a very small drop in the value of $C(\epsilon_2, \, v_2)$ (to 0.91). Whereas, $C(\epsilon_2, \, v_2)$ drops to a significantly lower value of 0.60 for 60--80\% centrality bin. It is to be noted that we have used a larger number of events for peripheral collisions to reduce the statistical uncertainty in the calculation. 

The correlation coefficient  $C(\epsilon_2, \, v_2)$ for kaon and proton  is also found to be large and positive (0.94 and 0.93 for kaon and proton respectively) for 0--20 and 20--40\% centrality bins. The value of C is 
 about 0.9 for 40--60\% and at  60--80\% centrality bin it drops significantly to a value of about 0.6 for both the particles.

\begin{figure}
\centerline{\includegraphics*[width=8.0 cm,clip=true]{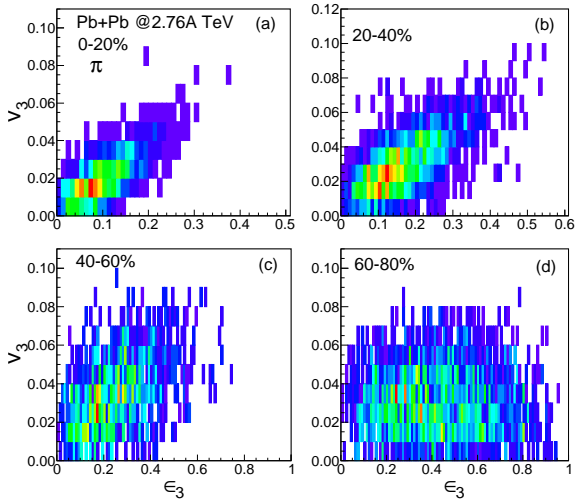}}
\caption{(Color online) Distribution of $\epsilon_3$ and $\pi^+$  $v_3$ at 2.76A TeV Pb+Pb collisions at four different centrality bins. }
\label{pi_v3}
\end{figure}

\begin{figure}
\centerline{\includegraphics*[width=8.0 cm,clip=true]{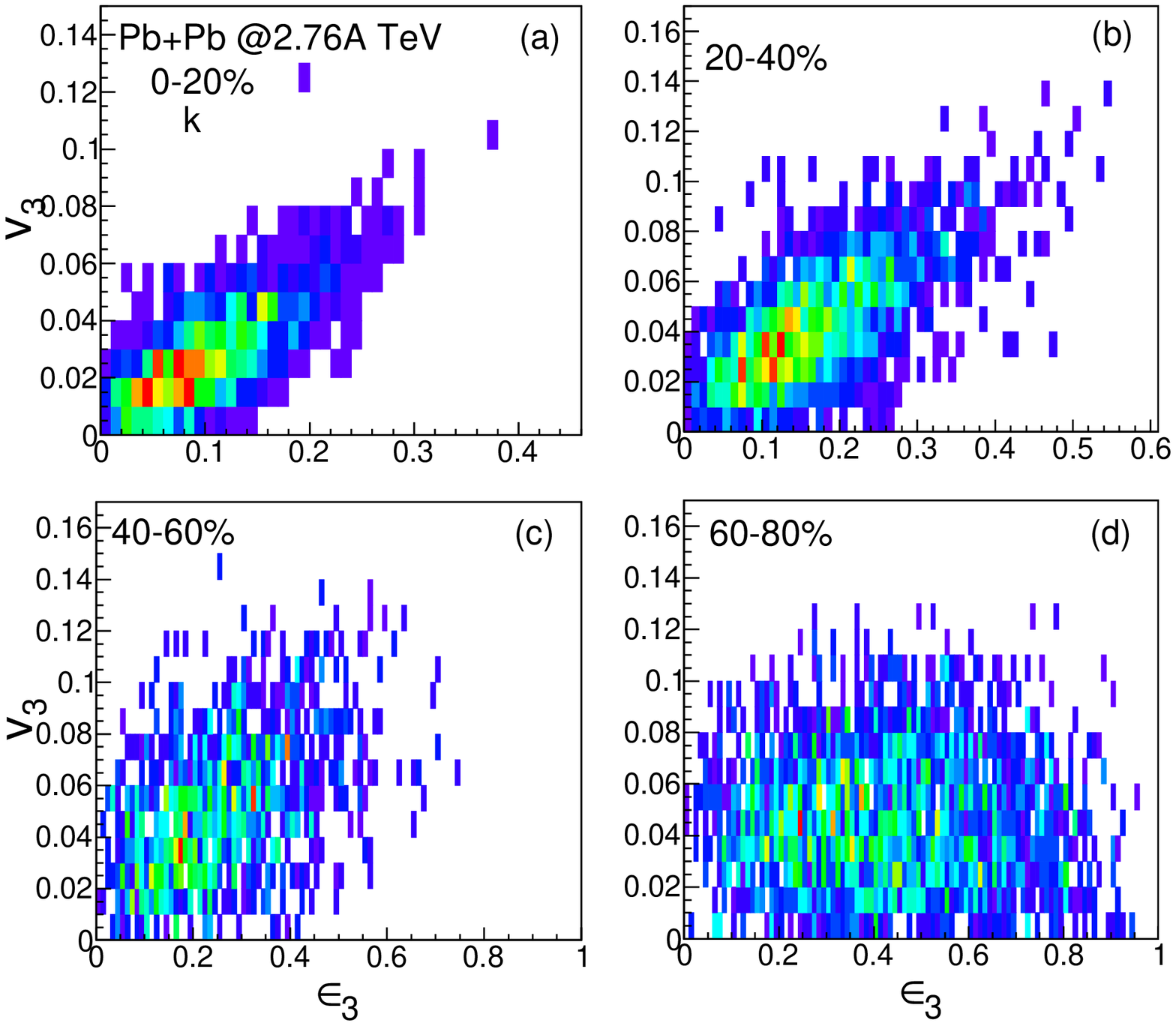}}
\caption{(Color online) Distribution of $\epsilon_3$ and $K^+$  $v_3$ at 2.76A TeV Pb+Pb collisions at four different centrality bins. }
\label{k_v3}
\end{figure}

\begin{figure}
\centerline{\includegraphics*[width=8.0 cm,clip=true]{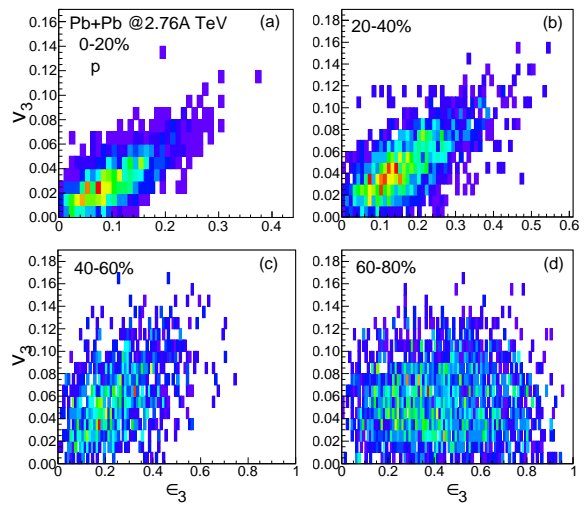}}
\caption{(Color online) Distribution of $\epsilon_3$ and $p$  $v_3$ at 2.76A TeV Pb+Pb collisions at four different centrality bins. }
\label{p_v3}
\end{figure}

It is well known that the triangular flow of charged particles does not show strong dependence on the collision centrality \cite{hannah}. However, the initial triangularity of the medium is found to be sensitive to the system size which makes the estimation of correlation coefficient between $\epsilon_3$ and $v_3$ important.  
The Figs. 4, 5, and 6 show the distribution of $\epsilon_3$ and $v_3$ for pion, kaon and protons respectively for the four different centrality bins at LHC. Many earlier studies have shown that the linear correlation between $\epsilon_3 - v_3$ is weaker than the correlation between $\epsilon_2 - v_2$~\cite{uli} and the same can be seen from the  figs (1-6) as well. 


The correlation coefficient C$(\epsilon_3, \, v_3)$ is about 0.75, 0.68 and 0.4 for 0-20\%, 20-40\% and 40-60\% centrality bins respectively for all $\pi^+$, $K^+$, and $p$ at LHC.
For 60--80\% centrality bin, we see  a complete absence of linear correlation as the value of C is found to be close to zero.

\begin{figure}
\centerline{\includegraphics*[width=8.0 cm,clip=true]{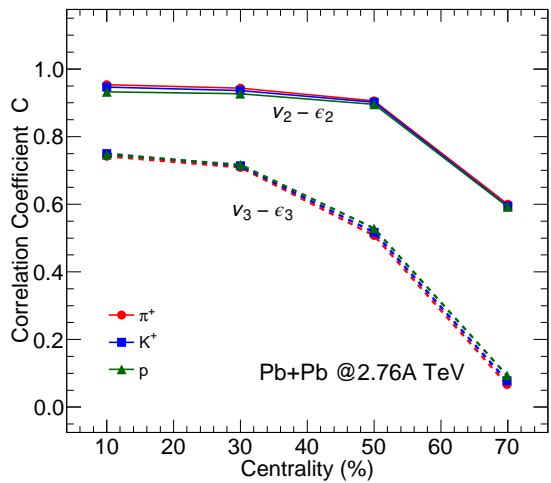}}
\caption{(Color online) C($\epsilon_n, v_n$) as a function of centrality from 2.76A TeV Pb+Pb collisions at LHC. }
\label{all_c}
\end{figure}

The correlation strength between $v_n$ and $\epsilon_n$ is summarized in the Fig.~\ref{all_c}. This figure clearly shows  the variation of correlation strength with collision centrality for the three hadrons. The correlation strength for $\pi^+$, $K^+$, and $p$ is found to be close to each other although the ($p_T$ integrated) anisotropic flow is different for them. We also see that  C($\epsilon_3,v_3$) shows a stronger sensitivity to the collision centrality as it decreases faster for peripheral collisions compared to C($\epsilon_2, v_2$).

We also estimate error in the correlation coefficient calculation with finite number of events.  The  probable errors in the correlation coefficients for ($\epsilon_2, v_2$) and ($\epsilon_3, v_3$) are found to be less than  1\% and 2\% respectively.

\section{correlation between $\epsilon_n$ and $v_n(p_T)$}
We understand that the final $p_T$ integrated anisotropic flow parameter is a consequence of the initial spatial deformation of medium formed in heavy ion collisions. Although the magnitude of both the elliptic flow and the spatial anisotropy increases while going away from central collisions, the correlation strength between $\epsilon_2$ and $v_2$ decreases towards peripheral collisions. We see similar reduction in correlation strength with collision centrality between $\epsilon_3$ and $v_3$ as well.

 The values of the correlation strength C($\epsilon_n, \ v_n$) for all $\pi^+$, $K^+$ and $p$ are found to be similar as a function of centrality and do not depend on the particle mass significantly as shown in Fig.~\ref{all_c}. We know that the mass ordering of differential anisotropic flow ($v_n(p_T)$) parameters is a signature of the collective behaviour of the medium formed in heavy ion collisions and it is well explained by hydrodynamical model calculation. Thus,  it is important to know if there also exists any mass dependence in the  $p_T$ dependent correlation coefficients for the different hadrons and the underlying mechanism does not depend on the particle mass significantly  which results in a similar correlation strength for them.

It is to be noted that the freeze-out temperature plays a crucial role in determining the $p_T$ dependent correlation coefficient considering our simplified assumption of constant temperature freeze-out for all hadrons. Some earlier studies have shown that the differential anisotropic flow parameter is sensitive to the freeze-out temperature mostly towards larger $p_T ( >1-1.5 \ {\rm GeV}) $ region~\cite{greco}.
A detail study using dynamical freeze-out conditions would be important, however the present study is also expected to provide valuable insight about the correlation between the initial geometry and the final  anisotropic flow parameters.

\begin{figure}
\centerline{\includegraphics*[width=8.0 cm,clip=true]{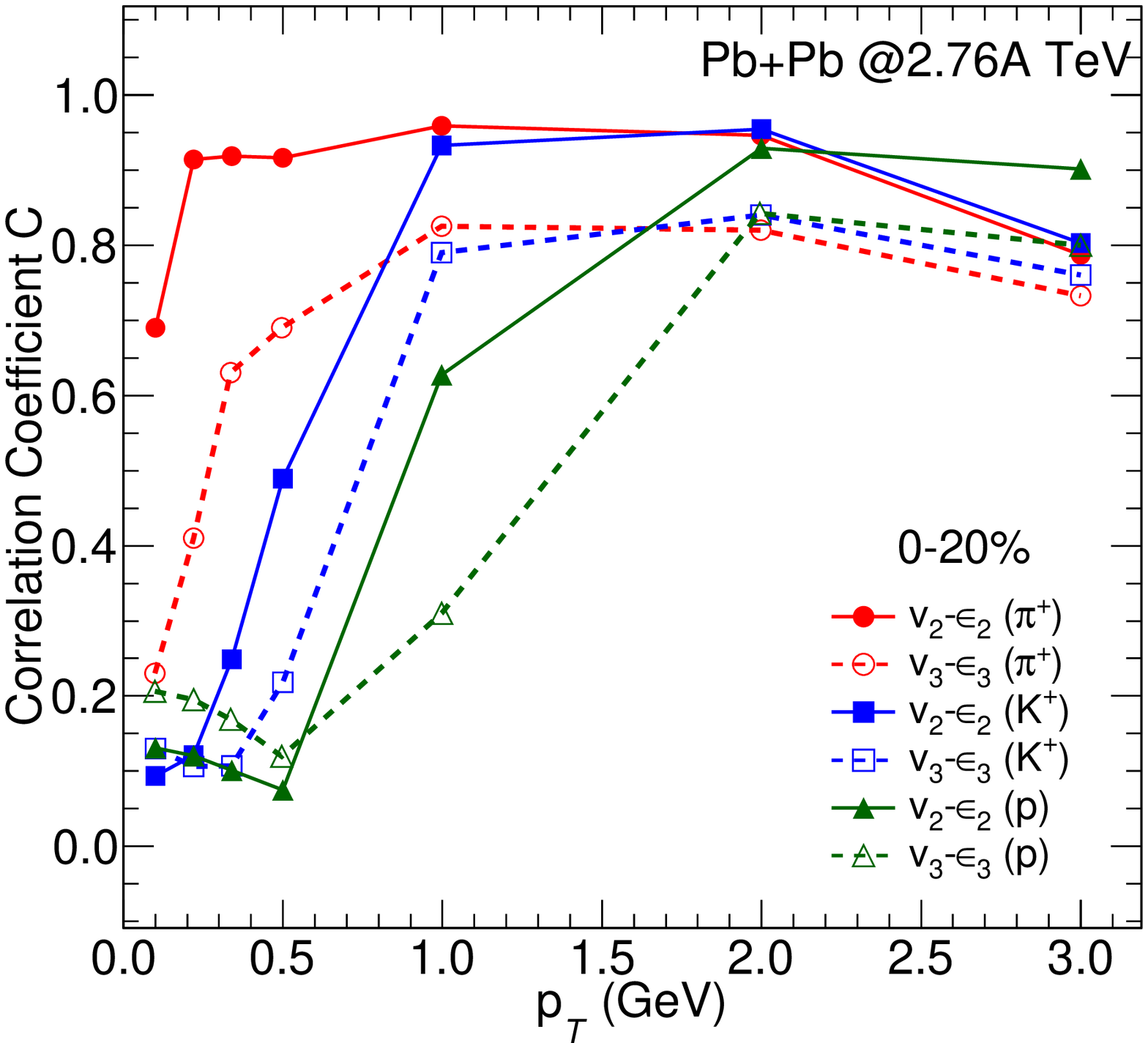}}
\caption{(Color online) (modified) Correlation between $\epsilon_n$ and $v_n(p_T)$ for pion at $\sqrt{s_{NN}}=$2.76 TeV Pb+Pb collisions for different centrality bins. }
\label{pi_v2_pt}
\end{figure}

\begin{figure}
\centerline{\includegraphics*[width=8.0 cm,clip=true]{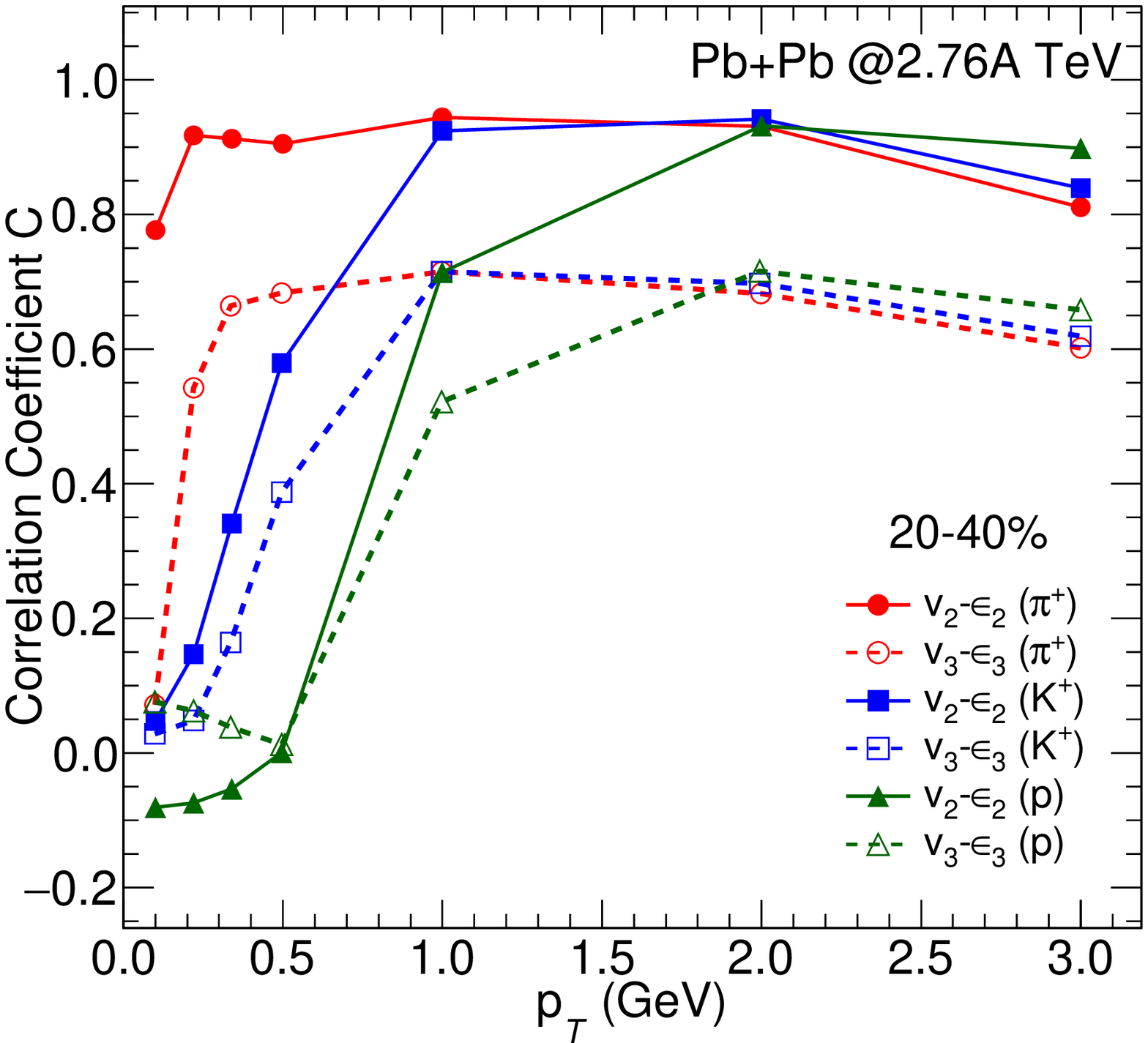}}
\caption{(Color online) (modified) Correlation between $\epsilon_n$ and $v_n(p_T)$ for kaon at $\sqrt{s_{NN}}=$ 2.76 TeV Pb+Pb collisions for different centrality bins. }
\label{k_v2_pt}
\end{figure}

\begin{figure}
\centerline{\includegraphics*[width=8.0 cm,clip=true]{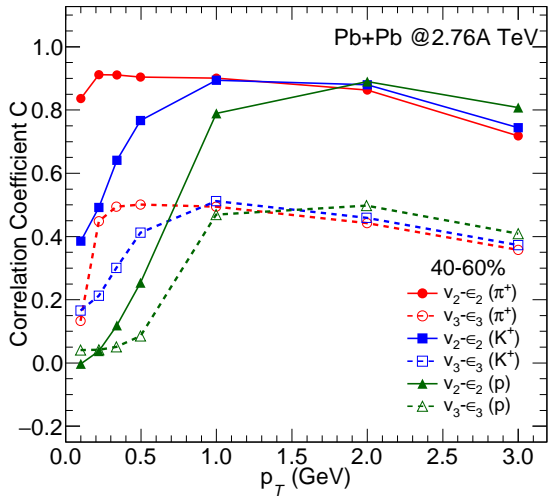}}
\caption{(Color online)(modified results) Correlation between $\epsilon_n$ and $v_n(p_T)$ for proton at $\sqrt{s_{NN}}=$ 2.76 TeV Pb+Pb collisions for different centrality bins. }
\label{p_v2_pt}
\end{figure}

The correlation coefficient C($\epsilon_n$, $v_n(p_T)$) for $\pi^+$, $K^+$, $p$ at different centrality bins  is shown in Figs~\ref{pi_v2_pt},~\ref{k_v2_pt}, and \ref{p_v2_pt} respectively.  A clear mass dependence in the correlation coefficient between $\epsilon_2 - v_2$ can be seen for all the centrality bins. 
The value of $C(\epsilon_2, v_2(p_T))$ is found to be larger for lighter particles in the   $p_T$ region 0.1 to 2 GeV shown in the figs. 

The strength of $\epsilon_2-v_2$ correlation  for pions remain close to 0.9 in the $p_T$ regain 0.2 to 2 GeV for all three centrality bins and then drops slowly for large $p_T$ values.  At very low $p_T$ ($< 0.3$ GeV), the value of C for $\pi^+$ is smaller, as those may be emitted from the initial few fm time period  when the build up of transverse flow is not very strong and the elliptic flow $v_2(p_T)$ is also small. 
We see a relatively stronger $p_T$ dependent correlation for $K^+$ and $p$  where the  coefficient  for them falls sharply with smaller $p_T$ values in the region $p_T < 1$ GeV. The correlation coefficient is found to be slightly larger for heavier particles in the region $p_T > $ 2 GeV.

The correlation  between $\epsilon_3-v_3$ as a function of $p_T$ also shows a similar behaviour to $\epsilon_2 - v_2$ although the magnitude is much smaller. The strength of correlation for protons is found to be very small below $p_T =$ 0.5 GeV for all centrality bins.  We see that C$(\epsilon_n, v_n(p_T))$  drops faster towards peripheral collisions for n=3 compared to n=2 at higher $p_T$ values.

These results clearly show that the $p_T$ dependent correlation coefficient strongly depends on the mass of the particle and $p_T$ region that contributes maximum to the correlation strength is also different for different particles. 
\begin{table}[h]
\centering
\begin{tabular}{|c|c|c|c|}
\hline
 (a) \ \ 0-20\%  \ & $\pi^+$   & $K^+$  & $p$ \\
\hline
C($\epsilon_2$, $v_2$) &  \ 0.95  \ & \ 0.95 \ & \ 0.93 \ \\
\hline
C($\epsilon_3$, $v_3$) & \ 0.74  \ & \ 0.75 \  & \ 0.75 \ \\
\hline

 (b) \ \ 20-40\%  \ &&& \  \\
\hline
C($\epsilon_2$, $v_2$) \ & \ 0.94 \ & \ 0.94 \ & \ 0.93 \ \\
\hline
C($\epsilon_3$, $v_3$) \ & \ 0.71 \ & \ 0.71 \ & \ 0.71 \ \\
\hline

 (c) \ \ 40-60\% \  &&& \ \\
\hline
C($\epsilon_2$, $v_2$) \ & \ 0.91 \ & \ 0.90 \ & \ 0.90 \ \\
\hline
C($\epsilon_3$, $v_3$) \ & \ 0.51 \ & \ 0.52 \ & \ 0.53 \ \\
\hline
 (d) \ \ 60-80\% \ &&& \  \\
\hline
C($\epsilon_2$, $v_2$) \ & \ 0.60 \ & \ 0.59 \ & \ 0.59 \ \\
\hline
C($\epsilon_3$, $v_3$) \ & \ 0.07 \ & \ 0.08 \ & \ 0.09  \ \\
\hline
\end{tabular}
\caption{{C($v_n,\varepsilon_n$) of $\pi^+$, $K^+$, and $p$ from (a) 0-20\%, (b) 20-40\%, (c) 40-60\%, and (d) 60-80\% Pb+Pb collisions at 2.76A TeV at the LHC .}}
\end{table}


\section{Cu+Cu collisions at RHIC}
The C$(\epsilon_n, v_n)$ of hadrons as a function of $p_T$ from Pb+Pb collisions at LHC is found to exhibit interesting  features which depend strongly on the mass of the particles. 
 Cu+Cu collisions at 200A GeV at RHIC are expected to produce a system with relatively smaller temperature and energy density as well as  smaller transverse dimension compared to Pb+Pb collisions at LHC. On the other hand, the initial state density fluctuations (increases anisotropic flow for smaller systems and lower beam energies) are expected to be higher for Cu+Cu collisions than for Pb+Pb collisions.  Thus, a comparison of the correlation coefficients from Cu+Cu and Pb+Pb systems at RHIC and LHC energies respectively is expected to provide better understanding of the beam energy and system size dependence of the correlation strength.

We study the correlation between $v_n$ and $\epsilon_n$ for three different centrality bins of Cu+Cu collisions at RHIC and compare with the results obtained from Pb+Pb collisions at LHC. Similar to the Pb+Pb collisions, the initial parameters are tuned to reproduce the experimental data of charged particle multiplicity and particle spectra for most central Cu+Cu collisions at RHIC. 

We consider $\tau_0$ as 0.4 fm/c and and $\eta$/s=0.08 for this case. A sufficiently large number of events have been generated for all 0--20\%, 20--40\%, and 40--60\% centrality bins of Cu+Cu collisions to calculate the $\epsilon_n$ and $v_n$ and the corresponding correlation coefficients between them.

\begin{figure}
\centerline{\includegraphics*[width=8.0 cm,clip=true]{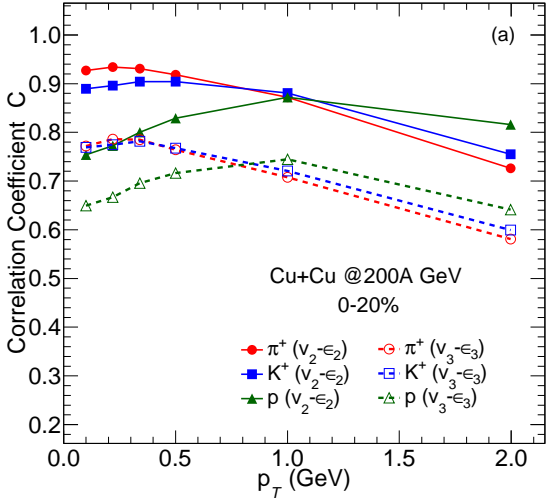}}
\centerline{\includegraphics*[width=8.0 cm,clip=true]{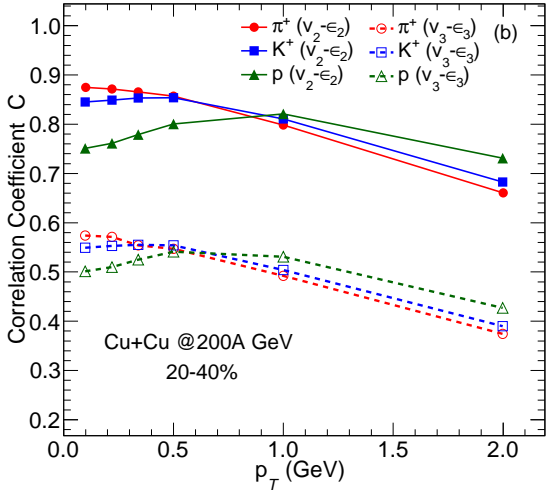}}
\centerline{\includegraphics*[width=8.0 cm,clip=true]{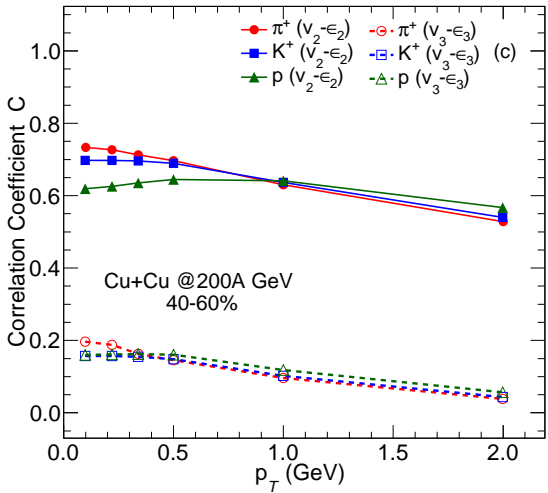}}
\caption{(Color online) Correlation coefficient C($\epsilon_n$,$v_n$) for $\pi^+$, $K^+$, and $p$ as function of $p_T$ for (a) 0--20\% (new result) and (b) 20--40\% centrality bins from Cu+Cu collisions at RHIC.}
\label{cucu}
\end{figure}

Again, for Cu+Cu collisions  we see that the correlation between $v_n$ and $\epsilon_n$ is relatively stronger for more central collisions and also for n=2 compared to n=3 (see Table II).

The Fig.~\ref{cucu} shows the $p_T$ dependent C($\epsilon_n, v_n$) for $\pi^+$, $K^+$ and $p$ for  different centrality bins. The C($\epsilon_n, v_n$) shows non-monotonic behavior as a function of $p_T$ specifically for protons.
We see a mass ordering of the correlation coefficient in a relatively narrower $p_T$ range for Cu+Cu collisions compared to Pb+Pb collisions.  For 0--20\% and 20-40\% centrality bins it can be seen in the range $p_T <$ 0.8 and 0.5 GeV respectively.  
For $\pi^+$ and $K^+$ the correlation strength as a function of $p_T$ is found to be maximum in the $p_T$ region 0.1 to 0.5 GeV and then it drops slowly for larger $p_T$ values. On the other hand, for heavier proton the strength of correlation is found to be  maximum around $p_T \sim$ 1 GeV. 

The anisotropic flow $v_n$ for Cu+Cu collisions is smaller than the same for Pb+Pb collisions for a particular centrality bin although the spatial anisotropy is slightly higher for the smaller system. In addition, the build up of transverse flow velocity is much weaker for Cu+Cu collisions and as a result the efficiency at which the spatial anisotropy is converted to momentum anisotropy is also relatively weaker for them compared to Pb+Pb collisions. 
Thus, for Pb+Pb collisions we see that even at 2 GeV $p_T$ value the correlation between ($\epsilon_n, v_n$) is still stronger whereas for Cu+Cu collisions the maximum contribution to the correlation strength comes from much smaller $p_T$ values.


These results show that the strength of correlation is larger for higher beam energies and is relatively weaker  for Cu+Cu collisions than for Pb+Pb collisions. 
The estimation of correlation coefficient from same type of system at different beam energies would provide more conclusive information about the dependence of C on particle mass and beam energy.

Fig. 12 shows the $\langle v_n \rangle / \langle \epsilon_n \rangle$ as a function of centrality for Pb+Pb collisions. Results from Cu+Cu collisions at RHIC for two centrality bins are also shown in the same plot for a comparison.
The slope ($C_n$) between two linearly correlated variables $ \epsilon_n $ and $ v_n$  can be written as  $  v_n \ = \ C_n \epsilon_n \ +\ \delta$.  After averaging over large number of events the slope is simply  $C_n \ = \ \langle v_n \rangle / \langle \epsilon_n \rangle$  as $\langle \delta \rangle$  is zero~\cite{Niemi}.  The $C_3$ for Pb+Pb collisions falls faster than $C_2$ towards peripheral collisions. 
The slope for Cu+Cu collisions is found to be much smaller than Pb+Pb collisions. However, the slope at peripheral Pb+Pb collision resembles the slope at the central Cu+Cu collisions, further assuring similar viscous effects from two different collision systems with multiplicities close to each other.  These results also clearly show that a smaller increase in spatial anisotropy results in a larger anisotropic flow for bigger system as well as for higher beam energies.

\begin{figure}
\centerline{\includegraphics*[width=8.0 cm,clip=true]{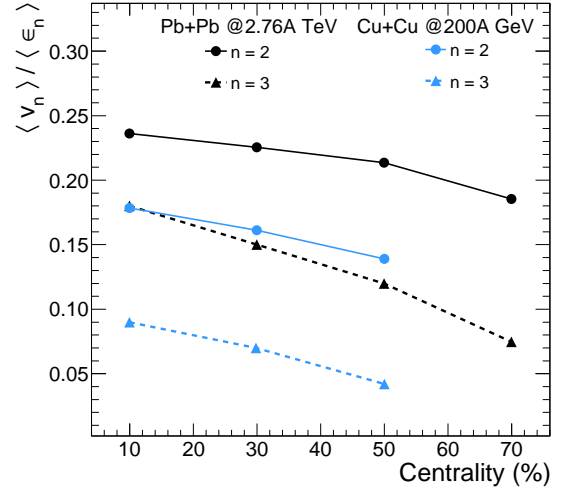}}
\caption{(Color online) The ratio of $\langle v_n \rangle$ and $\langle \epsilon_n \rangle$ as a function of collision centrality.}
\label{vn_en}
\end{figure}


\begin{table}[t]
\caption{C($\epsilon_n, v_n$) for 200A GeV Cu+Cu collisions at RHIC}
\begin{center}
\begin{tabular}{|c|c|c|c|}

\hline

 \ 0--20\%  \ & \  $\pi^+$ \ & \ $K^+$ \ & \ $p$ \ \\ \hline

C($ \epsilon_2$, $v_2$) \ & \ 0.91 \ & \ 0.90 \ & \ \ 0.90 \ \ \\ \hline

C($\epsilon_3$, $v_3$) \ & \ 0.72 \ & \  0.71 \ & \ \ 0.71 \ \ \\ \hline

20--40\%  &&& \ \\ \hline
 C($\epsilon_2$, $v_2$) \ & \ 0.84 \ & \ 0.83 \ & \ \ 0.82 \ \ \\ \hline

C($\epsilon_3$, $v_3$) \ & \ 0.51 \ & \ 0.50 \ & \ \ 0.50 \ \ \\ \hline

40--60\%  &&& \ \\ \hline
 C($\epsilon_2$, $v_2$) \ & \ 0.68 \ & \ 0.66 \ & \ \ 0.64 \ \ \\ \hline

C($\epsilon_3$, $v_3$) \ & \ 0.11 \ & \ 0.10 \ & \ \ 0.10 \ \ \\ \hline

\end{tabular}
\end{center}
\label{k_values}
\end{table}
%

\section{ $\eta/s$ dependence}
\begin{figure}
\centerline{\includegraphics*[width=8.0 cm,clip=true]{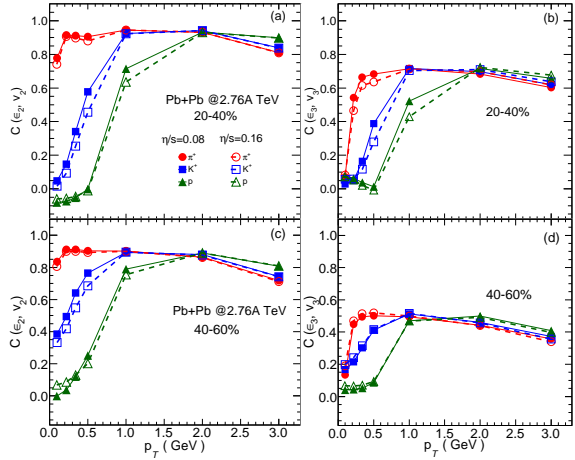}}
\caption{(Color online) $p_T$ dependent correlation coefficients at the LHC considering two different $\eta/s$ values.}
\label{eta_1}
\end{figure}

The the dependence of the $\epsilon_n-v_n$ correlation on the value of $\eta/s$ has been studied in detail in the literature. It has been shown in Ref.~\cite{Niemi} that the higher order correlation coefficients are more sensitive to the value of the $\eta/s$. In Fig~\ref{eta_1} we show the correlation coefficients for two different $\eta/s$ values for 20--40\% and 40--60\% Pb+Pb collisions at the LHC. The value of $C$ for all the three particles are found to vary only marginally when $\eta/s$ is changed from 0.08 to 0.16.

\begin{figure}
\centerline{\includegraphics*[width=8.0 cm,clip=true]{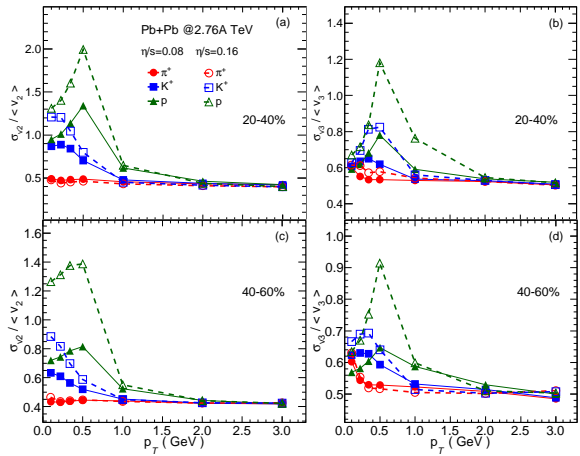}}
\caption{(Color online) Relative fluctuations in the anisotropic flow parameters at the LHC considering two different $\eta/s$ values.}
\label{eta_1.5}
\end{figure}

A better understanding of the initial state from final state flow observables has always been a primary goal for  flow analysis in heavy ion collisions. Due to the varying relation of the linear response parameter with multiplicity, it is challenging to relate the initial anisotropy to the final state momentum anisotropy in a linear fashion. The relative fluctuation in the anisotropic flow parameters  $\sigma_{v_n}/ \langle v_n \rangle$ is considered to be a potential observable, reflecting the ratio of the first two moments of the initial state eccentricity distribution (i.e, $\sigma_{\epsilon_n}$).
The relative fluctuations in the anisotropic flow parameters for the same set of collisions are shown in Fig.~\ref{eta_1.5}. Interestingly, the relative fluctuation $\sigma_{v_n}/\langle v_n \rangle$ as a function of $p_T$ is found to be quite sensitive to the value of $\eta/s$. One can see from the figures that the sensitivity to the value of $\eta/s$ is much stronger for protons than for pions and also in the low $p_T (< 1 {\rm GeV})$ region for Pb+Pb collisions. 

\section{Normalized symmetric cumulant}
The correlation between different order of anisotropic flow coefficients is studied in heavy ion experiments using cumulant method and  is considered to be an efficient method to reduce the non-flow effects in the measurements~\cite{nsc1,nsc2}.

The  Normalized Symmetric Cumulant  method has gathered a lot of attention in recent times which 
focuses on the correlation strength between different orders of anisotropic flow harmonics by removing the dependence on the magnitude of the harmonics.

We calculate NSC(2,3) between $v_2$ and $v_3$ using the relation,
\begin{equation}
{\rm {NSC}} (2,3) \, = \, \frac{\langle v_2^2 \, v_3^2\rangle \, - \, \langle v_2^2 \rangle \langle v_3^2 \rangle} {\langle v_2^2 \rangle \langle v_3^2 \rangle} \ .
\end{equation}

\begin{figure}
\centerline{\includegraphics*[width=8.0 cm,clip=true]{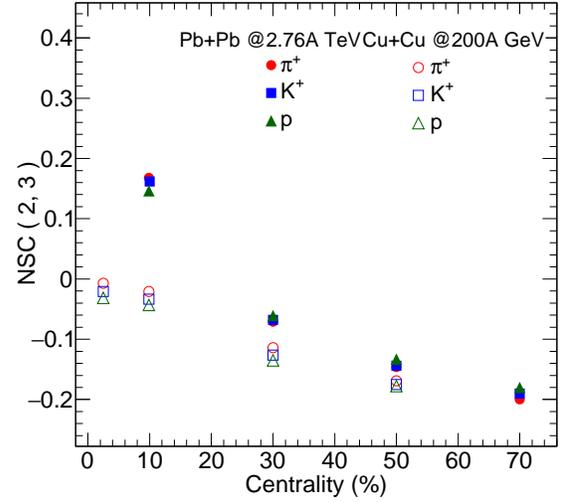}}
\caption{(Color online) NSC(2,3)  for pion, kaon and protons as a function of centrality from Pb+Pb collisions at LHC.}
\label{nsc}
\end{figure}

Fig.~\ref{nsc} shows the NSC between $v_2, v_3$  as a function of collision centrality for $\pi^+$, $K^+$ and $p$ from TeV Pb+Pb collisions at LHC and Cu+Cu collisions at RHIC. As expected, a clear anti-correlation between $v_2$ and $v_3$ can be observed for all the particles for collision centrality more than 20\%~\cite{prl}. Although the NSC(2,3) values are found to be close to each other for all $\pi^+$, $K^+$ and $p$ it is found to be slightly higher for heavier particles as a function of centrality. 

It is to be noted that the NSC(2,3) is found to be small and positive for Pb+Pb 0--20\% centrality bin and contrary to the peripheral collisions the value is found to be larger for lighter particles there. A similar observation  has been reported in earlier studies as well where it was shown that the centrality dependence of NSC(2,3) differs from most central to peripheral collisions~\cite{prl}.  
It has been shown in Ref.~\cite{nh7} that the value of NSC is sensitive to the size of the centrality bin specially for the central collisions. We see that the NSC  as a function of centrality does not change significantly with change in the value of $\eta/s$ from 0.08 to 0.16 for both Cu+Cu and Pb+Pb collisions.

\section{summary and Conclusions}
We calculate the  correlation between the initial spatial anisotropy  and the final momentum anisotropy for positively charged pion, kaon and proton from 2.76A TeV Pb+Pb collisions at LHC and at different centrality bins using an event-by-event viscous hydrodynamic model framework with fluctuating initial conditions.  The $\epsilon_n - v_n$ correlation from a relatively smaller system at lower beam energy (Cu+Cu collisions at 200A GeV at RHIC) is also calculated for a comparison with the Pb+Pb collisions.

The linear correlation  is found to be stronger for central collisions than for peripheral collisions for all the particles. In addition, the correlation between $\epsilon_3 - v_3$ is found to be weaker than $ \epsilon_2 - v_2$. However, the correlation between $v_n(p_T)$ and $\epsilon_n$ as a function of $p_T$ shows interesting behaviour where the correlation coefficient C is found to depend strongly on the mass of the particles. We see a clear ordering of the correlation coefficient in the lower $p_T$ region depending on the particle mass where the correlation strength is found to be larger for lighter particles. The $p_T$ range for the ordering depends on the collision centrality and also on the beam energy. The $p_T$ dependent correlation strength is found to rise with $p_T$, reach maximum, and then drop slowly beyond 2 GeV $p_T$ value for the Pb+Pb collisions. The correlation strength for $\pi^+$ reaches maximum at a relatively smaller $p_T$ value than for $K^+$ and protons. Although the strength of the correlation between $\epsilon_3$ and $v_3$ is found to be relatively weaker compared to $\epsilon_2$ and $v_2$, we see a similar qualitative $p_T$ dependent behaviour of correlation co-efficient for both of them.

The Cu+Cu collisions at RHIC produce a relatively smaller system than for Pb+Pb collisions for a particular centrality bin and the correlation coeffient C($\epsilon_n, v_n$) is found to be slightly smaller for Cu+Cu collisions. 
However, the $p_T$ dependent correlation coefficient from Cu+Cu collisions provide valuable insight about the system size and beam energy dependence of the correlation strength. The C($\epsilon_n, v_n (p_T)$) also  shows  mass ordering, however the $p_T$ range is much smaller (0.3 -- 0.8 GeV) compared to Pb+Pb collision. This could be due to a relatively weaker development of the transverse flow velocity for Cu+Cu collisions at lower beam energy than for Pb+Pb collisions at LHC. In addition, the correlation strength is found to be strongest at a much smaller $p_T$ value for Cu+Cu collisions than for Pb+Pb collisions. 
The study of correlation strength for same system at different beam energies would give a more quantitative estimation of the beam energy dependence.

The correlation coefficient is found to depend only marginally on the value of $\eta/s$. However, the relative fluctuations in the anisotropic flow parameter show strong sensitivity to the value of $\eta/s$. The value of $\sigma_{v_n}/\langle v_n \rangle$ is found to be significantly larger for larger $\eta/s$ for heavier particle and in the region $p_T < 1$  GeV.

We calculate the NSC(2,3) between the anisotropic flow coefficients for $\pi^+$, $K^+$, and $p$ for different centrality bins of Pb+Pb and Cu+Cu collisions. 
The NSC(2,3) as a function of collision centrality shows a clear anti-correlation between $v_2$ and $v_3$ for peripheral collisions and also does not show a strong dependence on the mass of the particles. 
\section{Acknowledgment}
We like to thank the VECC grid computing facility and Abhisekh Seal for the continuous help throughout this work. PD acknowledges the support of the National Natural Science Foundation of China under Grants No. 11835002 and No. 11961131011.

\end{document}